\documentclass[prl,reprint,sort&compress,aps,twocolumn,prb,superscriptaddress]{revtex4-2}
\usepackage{amssymb}
\usepackage{appendix}
\usepackage{verbatim}
\usepackage{float}
\usepackage{color}
\usepackage{textcomp}
\usepackage{gensymb}
\usepackage{hyperref}
\usepackage{dsfont} 
\hypersetup{
     colorlinks   = true,
     citecolor    = blue
}
\usepackage{graphicx}% Include figure files
\usepackage{dcolumn}% Align table columns on decimal point
\usepackage{bm}% bold math
\usepackage{braket}% bracket notation
\usepackage{empheq}
\usepackage{nicefrac} % for \nicefrac macro
\usepackage[dvipsnames]{xcolor}

\usepackage[normalem]{ulem}

\usepackage[utf8]{inputenc}
\usepackage{chngcntr}
\usepackage{xcolor}
\usepackage{braket}
\usepackage{physics}
\usepackage{sidecap}
\usepackage{lipsum}

\def\d{\downarrow}
\def\u{\uparrow}

\begin{document}

\title{
Strain induced quasi-unidimensional channels in twisted moir\'e lattices
}

\author{Andreas Sinner}
\email{andreas.sinner@uni.opole.pl}
\affiliation{IMDEA Nanoscience, Faraday 9, 28049 Madrid, Spain}
\affiliation{Institute of Physics, University of Opole, 45-052 Opole, Poland}

\author{Pierre A. Pantale\'on}
\email{ppantaleon@uabc.edu.mx}
\affiliation{IMDEA Nanoscience, Faraday 9, 28049 Madrid, Spain}

\author{Francisco Guinea}
\affiliation{IMDEA Nanoscience, Faraday 9, 28049 Madrid, Spain}
\affiliation{Donostia International Physics Center, Paseo Manuel de Lardiz\'abal 4, 20018 San Sebastián, Spain}
\affiliation{ Ikerbasque, Basque Foundation for Science, 48009 Bilbao, Spain}

\begin{abstract}

We study the effects of strain in moir\'e systems composed of honeycomb lattices. We elucidate the formation of almost perfect one-dimensional moir\'e patterns in twisted bilayer systems. The formation of such patterns is a consequence of an interplay between twist and strain which gives rise to a collapse of the reciprocal space unit cell. As a criterion for such collapse we find a simple relation between the two quantities and the material specific Poisson ratio. The induced one dimensional behavior is characterized by two, usually incommensurate, periodicities. Our results offer explanations for the complex patterns of one-dimensional channels observed in low angle twisted bilayer graphene systems and twisted bilayer dicalcogenides. %\textcolor{red}{ add something about the quasiperiodic bands!!}
Our findings can be applied to any hexagonal twisted moir\'e pattern and can be easily extended to other geometries.

\end{abstract}

\maketitle

%%%%%%%%%%%%%%%%%%%%%%%%%%%%%%%%%%%%%%%%%%%%%%%%%%%%%%%%%%%%%%%%%%%%%%%%%%%%%%%%%%%%%%%%%%%%%%%%%%%%%%%%

{\bf Introduction:} Twisted bilayer and multilayer systems represent two-dimensional materials, where atom-thick layers of the same or different materials are superimposed and rotated by an arbitrary twist angle. Twisted bilayer graphene (TBG) represents arguably the most prominent physical system of this kind~\cite{LopesdosSantos2007,Shallcross2008TwistB, Shallcross2010Turbo,Bistritzer2011,TramblydeLaissardire2010,Mele2010Comm}, the bilayers of transition metal dicalcogenides (TMD) the other~\cite{Xian2021MoS,Angeli2021Gamma,Naik2018UltraFlat,Wu2018TMD,Tang2020Hubb,Regan2020Mott,Wang2020Corre,Ni2019Soliton,Xian2019Multi}. The effect of twisting two periodic systems with respect to each other results in the formation of superlattices, the moir\'e patterns~\cite{Oster1963Moire,Oster1964Moire}. In TBG such moire patterns give rise to  very narrow bands at small twist angles, which can host correlated electronic states and superconductivity~\cite{Jarillo2018a,Jarillo2018b}. In addition, strains are ubiquitous in moiré systems~\cite{Metal21}. The interplay of electronic and elastic degrees of freedom in moiré systems is not fully understood~\cite{Kazmierczak2021Strain,deJong2022moire,Zhang2022Strain}.  
The effect of strains in monolayer graphene and other non-twisted bidimensional materials has been extensively studied~\cite{VKG2010,DWCF14,Aetal16,Naumis2017Review}. Important insights on the role of strains in twisted bilayer graphene were reported in~\cite{Huder2018Hetero,Fu2019}. 
The applied in-plane strain acting on both sublattices in opposite directions changes the distance between the nearest lattice atoms within each layer, and increases correspondingly the electronic hopping amplitude between them. In terms of the effective Dirac description of graphene, this effect creates an additional term which resembles the conventional vector potential, which however does not break the time-reversal symmetry of the Hamiltonian~\cite{VKG2010,Hasegawa2006,Wunsch2008,Pereira2009,Montambaux2009,Oliva2013Strain,OlivaLeyva2015Velo}.
This term displaces the Dirac points from their original positions but does not distort the shape of the Brillouin zone. This process breaks the $\bf C^{}_6$-symmetry of the Dirac points and lifts the degeneracy of the saddle points.
At larger strains the system goes through a Lifshitz transition characterized by a fusion of the Dirac points with resulting anysotropic spectrum and different scaling behavior of the low-energy part of the density of states~\cite{Hasegawa2006,Wunsch2008,Pereira2009,Montambaux2009,Oliva2013Strain,OlivaLeyva2015Velo}.

Similar effects might be expected for strained twisted bilayer graphene, such as the appearance of higher van-Hove singularities~\cite{Fu2019,Guinea2019,Pantaleon2022}. However the plethora of observed phenomena in strained twisted bilayer graphene is much larger than suggested by those analogies. For instance, the observation of highly anisotropic moir\'e patterns in the strained twisted bilayer graphene has been reported in many experiments~\cite{McEuen2013,Woods2021,Califer2021,Shabani2021,Ketal21,Jetal22,Jetal22b,Ketal22c}. With increasing strain the degree of deformation of the unit cells increases as well, until they become effectively one-dimensional stripes. 

In this work we show how the deformation of the moiré superlattice, and the emergence of quasi-one-dimensional features
is a consequence of the interplay between twist and strain.  As a criterion for such transition we find a simple relation between the applied uniaxial strain, the twist angle, and the material dependent Poisson ratio. 
%The effect of the formation of these pattern in the reciprocal space is the collapse of the Brillouin zone. 
Initially, the Brillouin zone has the form of a perfect honeycomb cell. With increasing strain it gets deformed and elongated in a selected direction, until it reduces to a line at the critical strain value. 
The selected direction is determined by the material dependent parameters. We construct the strain dependent lattice vectors in both real and reciprocal spaces and explore the consequences of this transition for the spectra and the density of states of twisted bilayer graphene within a continuum model approximation~\cite{LopesdosSantos2007,M10,Bistritzer2011,LopesdosSantos2012}.
In the one-dimensional limit we obtain electronic bands, which are determined by an interplay of two generally different and incommensurate periodicities. These can be fine-tuned to a single periodicity by varying external applied forces.

\begin{figure*}
\includegraphics[scale=0.66]{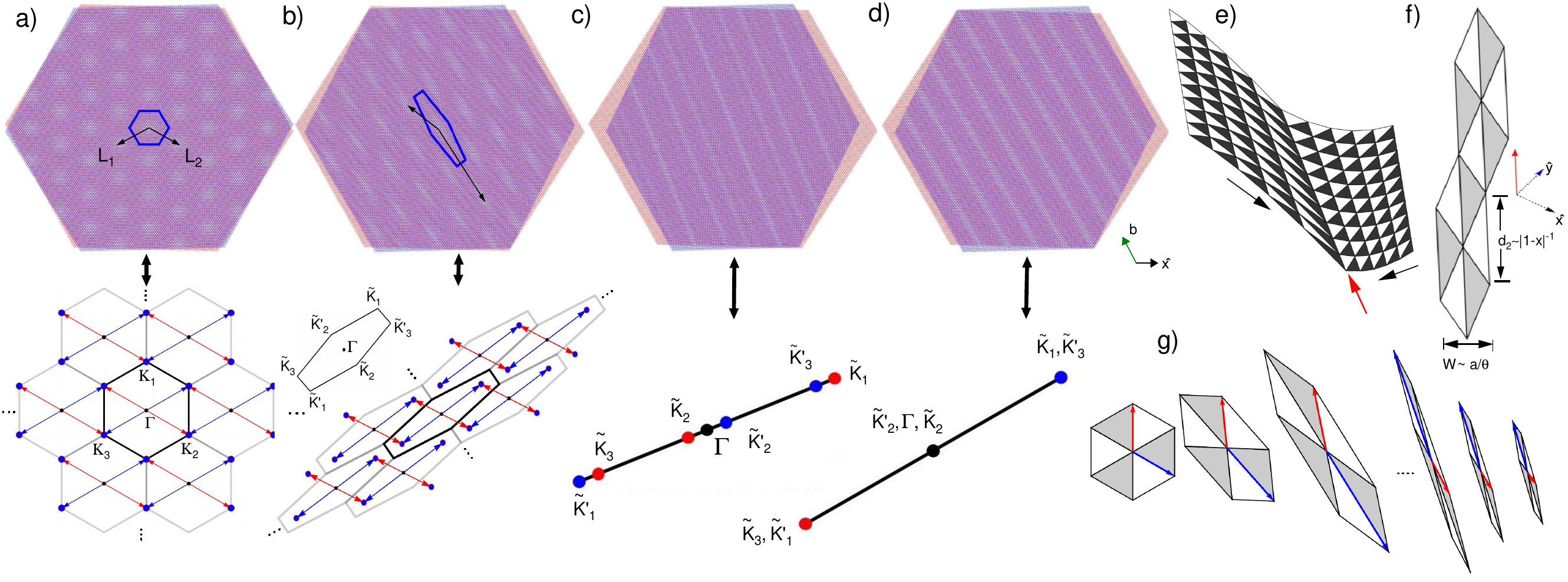}
\par
\caption{
Strain induced geometrical effects in a moir\'e superlattice with twist angle $\theta=3^\circ$ . The real space geometry of a twisted honeycomb bilayer subject to uniaxial heterostrain with Poisson ratio $\nu =0.165$ and strength: 
a) $\epsilon = 0\epsilon_{c}$, b) $\epsilon = 0.5 \epsilon_{c}$ and c) $\epsilon = 1.0 \epsilon_{c}$. In d) $\epsilon = 1.0 \epsilon_{c}$ and $\nu =1/3$, $\epsilon_{c}$ referring to as the critical strain value from Eq.~(\ref{eq:Collapse}). In a)-b) the blue hexagon visualizes the lattice unit cell. At the bottom of each moir\'e structure the corresponding reciprocal space is shown. Here, the colored arrows point to the geometrical positions of the Dirac points. In d) we show the formation of quasi-one dimensional channels due to a commensurate condition between the lattice vectors (see the main text). e) The formation of an edge domain wall, emphasized by the red arrow, due to a non-uniform strain. Bright (dark) triangles emphasize the AB (BA) stacking domains. Black arrows indicate the direction of the strain increasing from zero to a finite value, chose here as $0.5\epsilon_{c}$; f) Critical strain scales in real space for two unit cells. For visualization purposes, the hexagons have been rotated, as indicated by the $\hat{x}\hat{y}$ axes. g) Stacking domain cells calculated for $\epsilon=\{0, 0.25,0.50, 1.3, 1.4, 1.5\} \epsilon_{c}$,  from left to right. %For better visualization, all figures have been generated for the twist angle $\theta=3^\circ$
}
\label{fig:Figure1}
\end{figure*}

\vspace{3mm}
{\bf Geometry of the deformed moir\'e Brillouin zone:} We consider the case of a twisted bilayer honeycomb lattice, following the approach in~\cite{Fu2019}. We chose the reciprocal lattice vectors for each monolayer system as 
${\bf b}^{}_{1,2}={2\pi}{a}^{-1} \left(1, \mp{1}/{\sqrt{3}} \right)$, with lattice constant $a$ (for graphene $a\approx 2.46 $ \AA). The reciprocal lattice vectors of the twisted layers are obtained by rotating vectors ${\bf b}^{}_i$ by a twist angle, ${\bf G}^{\u,\d}_i = {\rm R}\left[\mp\theta/2\right]{\bf b}^{}_i$,  $\u/\d$  (and respectively -/+ at the twist angle) referring to the upper/lower layer, ${\rm R}[\pm\theta]$ being the usual rotation matrix, cf. Eq.~(\ref{eq:RotMat}) in the Supplement~\cite{SI}.  
The reciprocal lattice vectors of the  moir\'e superlattice are ${\bf g}^{}_i = {\bf G}^{\u}_i - {\bf G}^{\d}_i$. 
Being subject to the geometric deformation by strain they change to 
\begin{eqnarray}
\label{eq:StrBasis}
\tilde {\bf g}^{}_i  = \tilde{\bf G}^{\u}_i - \tilde{\bf G}^{\d}_i = {\bf T}{\bf b}^{}_i ,
\label{reciprocal}
\end{eqnarray}
where the transformation matrix ${\bf T}$ is:
\begin{equation}
\label{eq:TransMatr} 
{\bf T} = \left(\mathds{1} - {\cal E}^\u\right) {\rm R}\left[-\theta/2\right]
 - \left(\mathds{1} - {\cal E}^\d\right){\rm R}\left[+\theta/2\right],
\end{equation}
with the symmetric strain tensor ${\cal E}^{\ell}=\{\epsilon^{\ell}_{i,j}\}$, $i,j=\{x,y\}$ and $\ell=\u,\d$. In the experimentally relevant case of uniaxial heterostrain, in which forces are applied along one direction
one makes a simplification ${\cal E}^\d = - {\cal E}^\u = {\cal E}/2$~\cite{Huder2018Hetero}. In particular, the uniaxial heterostrain can be parameterized in terms of two quantities: the dimensionless 
strain magnitude $\epsilon$ measured with respect to the lattice spacing,
and the strain direction determined by the angle $\phi$. In this case the strain tensor ${\cal E}$ becomes
\begin{equation}
\label{eq:uhs}
{\cal E} = 
\left(
\begin{array}{cc}
 \epsilon(\nu\sin^2\phi - \cos^2\phi) & \epsilon(1+\nu)\sin\phi\cos\phi \\
 \epsilon(1+\nu)\sin\phi\cos\phi      & \epsilon(\nu\cos^2\phi - \sin^2\phi) 
\end{array}
\right),
\end{equation}
where $\nu$ is the Poisson ratio of the system's monolayers. For monolayer graphene this value is roughly $\nu \approx 0.16$. Here we emphasize that our considerations include but are not restricted to the only case of the honeycomb uniaxial heterostrain. Note that when forces along two perpendicular directions are applied, the strains $\epsilon_{\parallel,\parallel}$ and $\epsilon_{\perp,\perp}$ can be tuned separately.

\begin{figure*}
\includegraphics[scale=0.71]{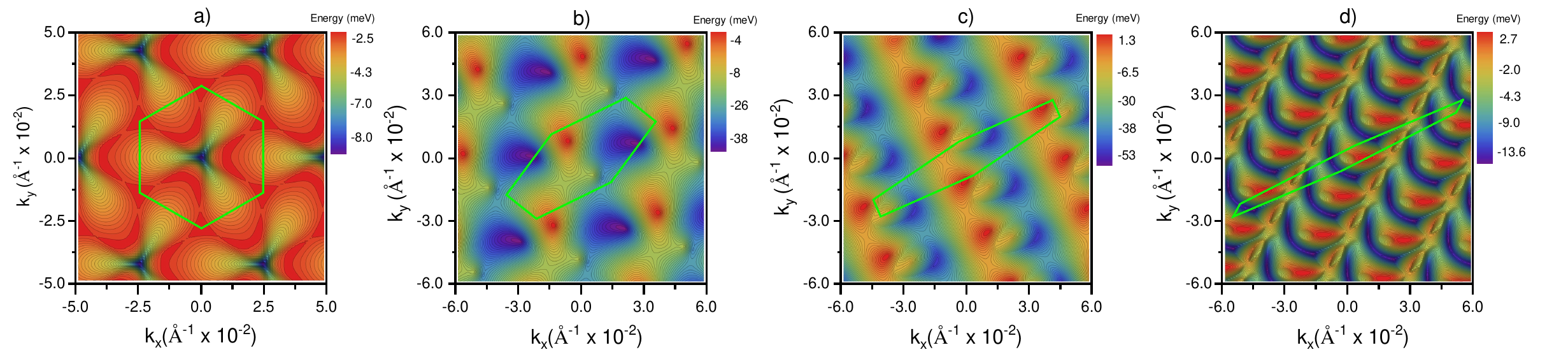}
\par
\caption{
The evolution of the bottom middle band of TBG with the twist angle $\theta=1^\circ$ as a function of uniaxial heterostrain with Poisson ratio $\nu = 0.165$ and: a) $\epsilon = 0~\epsilon_{c}$, b) $\epsilon = 0.3~\epsilon_{c}$, c) $\epsilon = 0.60~\epsilon_{c}$ and d) $\epsilon = 0.80~\epsilon_{c}$, with $\epsilon_{c}$ denoting the critical strain value defined in Eq.~(\ref{eq:Collapse}). Green hexagons highlight the boundaries of the moir{\'e} Brillouin zone.}
\label{fig:Figure2}
\end{figure*}

\vspace{3mm}
{\bf Collapse of the moir\'e Brillouin zone for a critical value of the strain:} 
In Fig.~\ref{fig:Figure1} we show the real and reciprocal space of a twisted moir\'e lattice. As the strain increases, the unit cell in real space is enlarged and rotated with tendency towards a particular spatial direction. In reciprocal space, the Brillouin zone gets progressively squeezed until it collapses at a certain critical value. This collapse implies that the vectors ${\tilde{\bf g}}_i$ in Eq.~(\ref{reciprocal}) are collinear, which occurs when
$
{\bf T}{\bf b}^{}_i = \alpha^{}_i{\bf s}, i=1,2
$
for  vector $\bf s$ to be determined and real $\alpha^{}_i$. Combining the equations we get 
\begin{eqnarray}
{\bf T}\left[{\bf b}^{}_1 - \frac{\alpha^{}_1}{\alpha^{}_2}{\bf b}^{}_2\right] = {\bf 0} . 
\end{eqnarray}
Since ${\bf b}^{}_i$ are linearly independent, %i.e. there is no such number $\alpha$ that $\alpha{\bf b}^{}_i ={\bf b}^{}_j\;\; {\rm for }\;\; i\neq j$, 
the above equation is satisfied only if 
\begin{equation}
\label{eq:CollapseEq}
\det{\bf T} = 0. 
\end{equation}
Equation~(\ref{eq:CollapseEq}) represents one of our main results. It is very general and does not assume any specific lattice, nor any specific type of the strain. For the particular case of the uniaxial heterostrain Eq.~(\ref{eq:uhs}), the collapse condition, Eq.~(\ref{eq:CollapseEq}), reduces to

\begin{equation}
\label{eq:Collapse}
\epsilon^{}_c = \pm \frac{2}{\sqrt{\nu}} \tan\frac{\theta}{2},
\end{equation}
which is independent angle $\phi$, and $\epsilon^{}_c$ denotes the critical strain strength. The linearization of Eq.~(\ref{eq:Collapse}
) $\epsilon^{}_c\approx\pm\theta/\sqrt{\nu}$ is reasonable for $\theta<10^\circ$. This result implies, that for a small angle moir\'e lattice the critical strain should be within the experimental range. For example, for TBG with a marginal twist angle $\theta<0.2^\circ$, an uniaxial heterostrain lower than $1.0$ \% is required to collapse the Brillouin zone and this create unidimensional channels. Importantly, this is consistent with previous observations of unidimensional domains in bilayer graphene systems~\cite{McEuen2013}.    

{\bf Geometry of the deformed honeycomb moir\'e lattice:} We now explore the consequences of the collapsing conditions obtained in the previous section. For simplicity but without loss of generality we assume that the strain direction is along the $x$ axis, i.e. for $\phi=0$. The six corners of the zigzag oriented monolayer honeycomb Brillouin zone
\begin{equation}
{\bf K}^{}_{1}  = k^{}_0
\left( 
\begin{array}{c}
 -2 \\ 0 
\end{array}
\right), \;\;
 {\bf K}^{}_{2,3} = k^{}_0
 \left( 
 \begin{array}{c}
  1 \\ \displaystyle \pm\sqrt{3}
 \end{array}
 \right), 
\end{equation}
and  ${\bf K}^\prime_{1} = - {\bf K}^{}_{3}$, ${\bf K}^\prime_2 =-  {\bf K}^{}_1 $, and ${\bf K}^\prime_3 = - {\bf K}^{}_2$, 
where $k^{}_0={2\pi}(3a)^{-1}$. The six corners of the deformed armchair oriented moir\'e Brillouin zone via
\begin{equation}
\label{eq:Kvectors}
\tilde{\bf K}^{}_i = {\bf T} {\bf K}^{}_i.
\end{equation}
Expressing the strain strength in terms of its critical value as $\epsilon = x\epsilon^{}_c$ where we introduce a strain parameter $0<x<1$, cf. Fig.~\ref{fig:Figure1} 

\begin{eqnarray}
\tilde {\bf K}^{}_{1} &=& \tilde k^{}_0
\left(\begin{array}{c}
\frac{2 x}{\sqrt{\nu}} \\ 2 
\end{array} \nonumber
\right),  \\
\tilde {\bf K}^{}_{2,3} &=&{\tilde k}^{}_0
\left( 
\begin{array}{c}
 \pm \sqrt{3} - \frac{x}{\sqrt{\nu}} \\  
  -1 \pm \sqrt{3\nu}x  
\end{array}
\right),
\label{eq:rvectors}
\end{eqnarray}
where $\tilde k_0= 2k_0\sin[\theta/2]$. At the critical strain $x=1$ all these vectors are multiples of 
\begin{equation}
%\label{eq:s-vec}
{\bf s} =  \left( 
\begin{array}{c}
 \frac{1}{\sqrt{\nu}} \\  1 
\end{array}
\right),
\label{bzdirection}
\end{equation}
which turns out to depend only on the material specific Poisson ratio and explains the selected direction in the momentum space clearly visible in Figs.~\ref{fig:Figure1} and~\ref{fig:Figure2}.

\begin{figure*}
\includegraphics[scale=0.50]{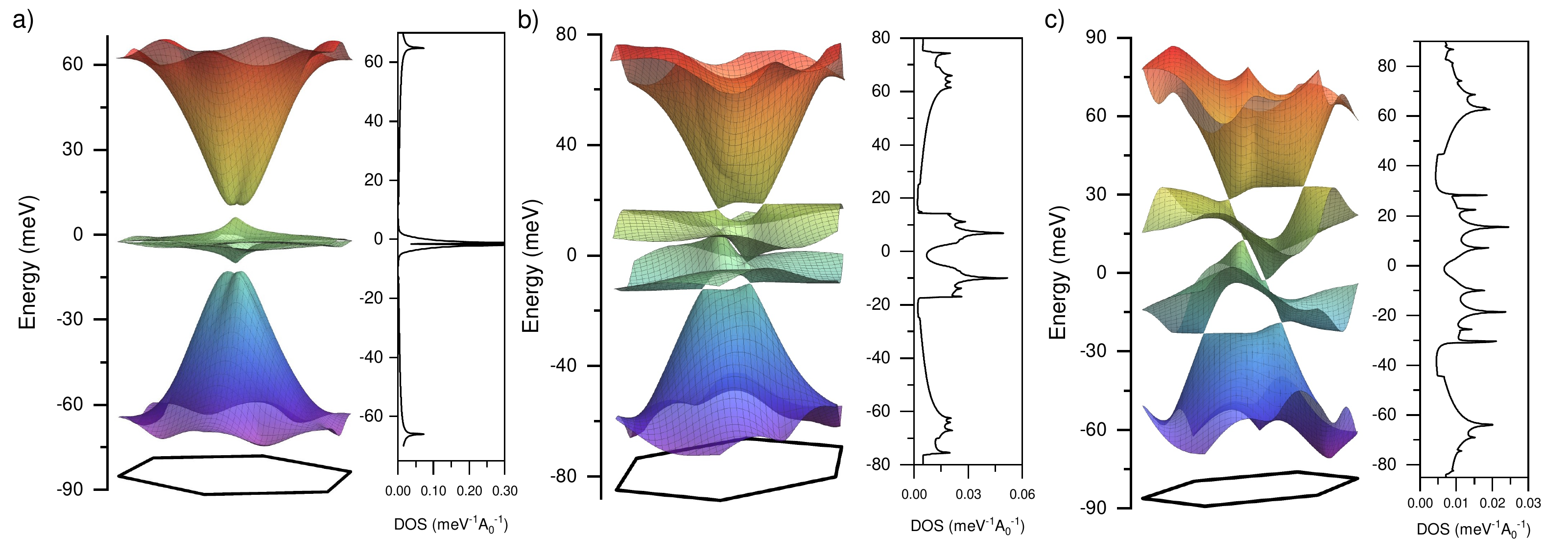}
\par
\caption{
Evolution of the band structure and DOS of TBG with the twist angle $\theta=1.0^\circ$ as a function of the uniaxial heterostrain with Poisson ratio $\nu = 0.165$ and: a) $\epsilon = 0~\epsilon_{c}$, b) $\epsilon = 0.10~\epsilon_{c}$ and c) $\epsilon = 0.20~\epsilon_{c}$. The corresponding moir{\'e} Brillouin zone is shown underneath each spectrum.}
\label{fig:Figure3}
\end{figure*}

The real space unit cell is oriented along a direction perpendicular to the vector in Eq.~(\ref{bzdirection}),
\begin{equation}
{\bf b} =  \left( 
\begin{array}{c}
-\sqrt{\nu} \\ 1 
\end{array}
\right).
\label{ucdirection}
\end{equation}
Equations~(\ref{eq:CollapseEq}), (\ref{eq:Collapse}), (\ref{bzdirection}), and (\ref{ucdirection}) represent central results of our work and they are valid for any twisted honeycomb lattice system.

We chose the reciprocal lattice vectors of the moir\'e superlattice to be
\begin{equation}
\label{eq:LatVec}
\tilde {\bf g}^{}_1 = \tilde {\bf K}^{}_3 - \tilde{\bf K}^{}_1, \;\; \tilde {\bf g}^{}_2 = \tilde{\bf K}^{}_2 - \tilde {\bf K}^{}_1, 
\end{equation}
and finally obtain the real space lattice vectors of the distorted lattice in the form: 
\begin{eqnarray}
\label{eq:RealSpaceV1}
{\bf L}^{}_1 & = & \frac{\sqrt{3}a}{4\sin[\theta/2]}\frac{1}{1-x^2}
\left( 
\begin{array}{c}
 - \sqrt{3}+x\sqrt{\nu} \\ -1+x\sqrt{\frac{3}{\nu}} 
\end{array}
\right), \\ 
\label{eq:RealSpaceV2}
{\bf L}^{}_2 & = & \frac{\sqrt{3}a}{4\sin[\theta/2]}\frac{1}{1-x^2}
\left( 
\begin{array}{c}
 \sqrt{3}+x\sqrt{\nu} \\ -1-x\sqrt{\frac{3}{\nu}} 
\end{array}
\right).
\end{eqnarray}
Note that the length of the vectors ${\bf  L^{}_{1}} , {\bf  L^{}_{2}}$ diverges at the critical strain, $x \rightarrow 1$ as $1/| 1-x|$. Below we assume the twist angle $\theta$ to be small and keep it to the leading order only. The area of the real space unit cell is 
\begin{align}
A_{0} &= | {\bf L^{}_{1}} \times {\bf L^{}_{2}} | = \frac{3 \sqrt{3}a^2}{2\theta^2| 1 - x^2|} \sim \frac{1}{| 1 - x |}
\end{align}
close to the critical strain value. 
We define the length of the unit cell as 
\begin{align}
    L &= \frac{| {\bf L{}_{1}} + {\bf L^{}_{2}}|}{2} = \frac{3 a \sqrt{1 + \nu x^2}}{2 \theta ( 1 - x^2)} \sim \frac{1}{| 1 - x |}
\end{align}
and the width of the unit cell as:
\begin{align}
    W &= \frac{\left|  {\bf L^{}_{1}}
 \times ( {\bf L^{}_{1}} + {\bf L^{}_{2}} ) \right|}{| {\bf L^{}_{1}} + {\bf L^{}_{2}} |} = \frac{3a}{2\theta \sqrt{1 + \nu x^2  }} 
 \end{align}
The width of the unit cell remains finite at the critical strain, $x \rightarrow 1$.

Near the critical strain, $x \rightarrow 1$, the modulations of the moiré lattice in the direction normal to the unit cell, Eq.~(\ref{bzdirection}) are determined by the reciprocal lattice vectors, in Eq.~(\ref{eq:rvectors}).  The lengths of these vectors are proportional to $1 -\sqrt{3 \nu}$ and $1 + \sqrt{3 \nu}$. The ratio between these values is, generally, incommensurate. Hence, the properties of the material, near the critical strain, are determined by a unit cell which diverges in one direction, and by combinations of non commensurate periodicities in the other direction. 

\begin{figure*}[t]
\includegraphics[scale=0.55]{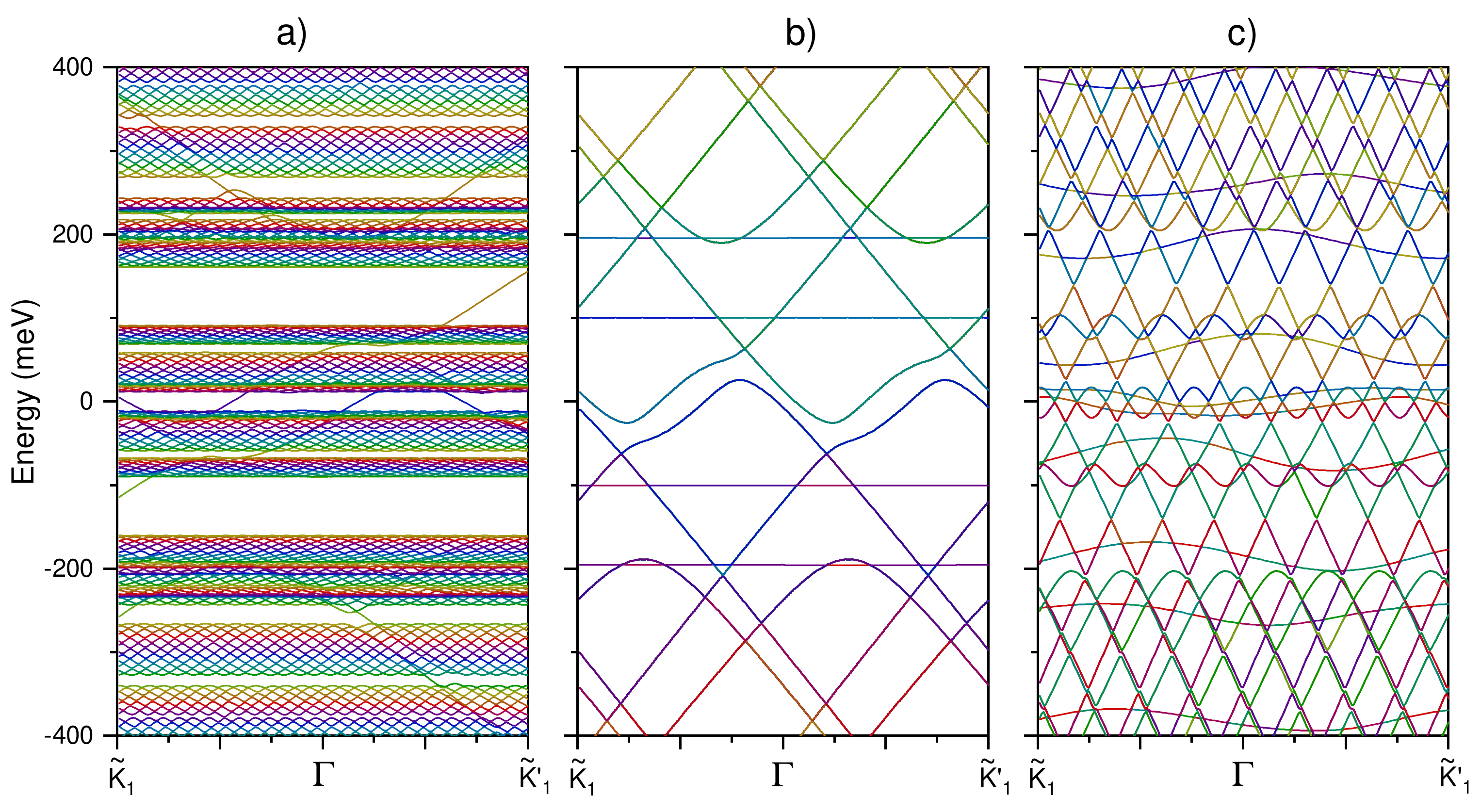}
\par
\caption{
Band structures of the critically strained TBG for $\theta = 1^{\circ}$ and different values of the ratio between applied strains: a) $- \epsilon_{yy} / \epsilon_{xx}$ = 0.165, b) $- \epsilon_{yy} / \epsilon_{xx}$ = 1/3 and c) $- \epsilon_{yy} / \epsilon_{xx}$ = 1/12. The spectra are evaluated along the collapsed Brillouin zone depicted in Fig.~\ref{fig:Figure1}. }
\label{fig:Figure4}
\end{figure*}

{\bf Other strain combinations:}
The analysis so far has been restricted to uniaxial heterostrains, induced by forces of opposite sign at both layers and at both ends of the sample. Other forces applied at the boundaries can lead to different patterns of strains inside the system. When normal forces are applied to boundaries rotated by $90^\circ$, the relation $\epsilon_{yy} / \epsilon_{xx}$ is no longer fixed by the Poisson ratio $\nu$. The pattern is simplified when the periodicities in Eq.~(\ref{eq:rvectors}) are commensurate. This happens when:
\begin{align}
    \frac{1 - \sqrt{- \frac{3 \epsilon_{yy}}{\epsilon_{xx}}}}{1 + \sqrt{- \frac{3 \epsilon_{yy}}{\epsilon_{xx}}}} &= \frac{m}{n}
    \label{periodicities}
\end{align}
where $m$ and $n$ are integers. This equation is satisfied when:
\begin{align}
    - \frac{\epsilon_{yy}}{\epsilon_{xx}} &= \frac{( m - n )^2}{3 ( m + n )^2}
    \label{eq:PoisonInte}
\end{align}
The simplest solutions are $m= 0 , n = 1$, giving $- \epsilon_{yy} / \epsilon_{xx} = 1 / 3$, and $m= 3 , n = 1$, giving $- \epsilon_{yy} / \epsilon_{xx} = 1 / 12$. Then, the properties of the system are determined by a single periodicity, and the pattern at the critical strain reduces to a one dimensional lattice of $AA, AB$ and $BA$ stripes as shown in Fig.~\ref{fig:Figure1}c) and d) (see also  Fig.~\ref{fig:RealCritical} in the Supplement~\cite{SI}).

\vspace{5mm}
{\bf Non-homogeneous strain distributions:}
As the size of the moiré unit cell near the critical strain diverges, small local variations of the strain can lead to large changes of the moiré pattern. This is consistent with several recent experimental studies, where the creation of different types of moir\'e lattice defects has been reported. These are for instance the domain walls between different stacking domains in TBG~\cite{McEuen2013},  hexagonal boron nitride~\cite{Woods2021}, or transition metal dichalcogenides~\cite{Shabani2021}. In Fig.~\ref{fig:Figure1}d) we show a strain induced inhomogeneous moir\'e pattern, where the strain increases in the direction indicated by the black arrows. Here, the moir\'e pattern interpolates continuously between the minimal and maximal strain. Red arrow points to the region, where the formation of the edge domain wall occurs. This particular inhomogeneous pattern is constructed by a linear combination of the lattice vectors defined in Eqs.~(\ref{eq:RealSpaceV1}) and~(\ref{eq:RealSpaceV2}). Such inhomogeneous patterns have a strong resemblance to those reported in Refs.~\cite{Shabani2021,Woods2021}. 
Further realizations of deformed systems for various strains can be constructed from distorted hexagons, as for instance those shown in Fig.~\ref{fig:Figure1}g). More examples are presented the Supplement~\cite{SI}. It is worth noting that inhomogeneites at the atomic scale are magnified in moiré superlattices~\cite{DWCF14,PM20,Metal22}. We leave for later work the study of the interplay between atomic defects, twists, and strains.

\vspace{2mm}
{\bf Electronic spectrum and the density of states of twisted bilayer graphene with strain:}
We now turn our attention to TBG and focus on the twist angle near the first magical angle, $\theta = 1^{\circ}$. Because of the vector potential defined in Eq.~(\ref{eq:VectPot}) in the Supplement~\cite{SI}, in a strained TBG the Dirac points do no longer reside at the corners of the moir\'e Brillouin zone, cf. Fig.~\ref{fig:Figure2}. As strain increases, the Dirac points of the middle bottom band shown in Fig.~\ref{fig:Figure2} move on an involved trajectory within the Brillouin zone, which is influenced by several factors, e.g. the geometry of the deformed moiré Brillouin zone and interlayer coupling between graphene layers. More details can be found in the Supplement~\cite{SI}. The strain reduces the ${\bf C}^{}_3$-symmetry of the TBG moir\'e superlattice to the mere  ${\bf C}^{}_2$-symmetry, cf. Fig.~\ref{fig:Figure1}. This reduction in the symmetry lifts all spectral degeneracies protected by the ${\bf C}^{}_3$-symmetry, such as the positions of nodal and saddle points in the middle bands and similar spectral features of the remote bands, which are shifted with respect to each other on the energy axis. 
In the process of approaching the critical strain, the bands of the initially two-dimensional system are gradually deformed and compressed down to those of some effective one-dimensional system.

The analysis of the middle bands under strain reveals an intricate dynamics of their saddle points, cf. Fig.~\ref{fig:Figure2} and~\ref{fig:Figure3}. 
At zero strain all three saddle points are located at the same energy, i.e. there is a three-fold saddle point degeneracy. This degeneracy is protected by the $\bf C^{}_3$-symmetry between the saddle points in each of the bands, which is broken by the strain. This leads to the loss of the saddle point degeneracy and gives rise to a multiple peaked structure in the DOS, cf. Fig.~\ref{fig:Figure3}b) and c).

{\bf Electronic properties of the TBG's continuum model at the critical strain:} 
As mentioned earlier, at the critical strain the electronic wavefunctions are determined by the competition between the two, usually incommensurate, periodicities shown in Eq.~(\ref{periodicities}) (see also Eq.~(\ref{eq:periodcond}) in the Supplement~\cite{SI}). The resulting equations resemble the Harper equation~\cite{H55}, extensively discussed in connection to lattice electrons in a constant magnetic field. Systems described by variations of the Harper's equation typically show a discontinuous density of states, and either localized or extended states~\cite{S81,S81b,TN83,NR86,KKL86,TM20,TM21}. We present results for the electronic states for commensurate and incommensurate periodicities, and a twist angle $\theta = 1^\circ$, in Fig.~\ref{fig:Figure4}, cf. also Fig.~\ref{fig:ComparisonCritical} in the Supplement~\cite{SI}. The bands are plotted in a Brillouin zone defined by the sum of the two periodicities. For incommensurate combinations, the results are consistent with extended, i.e. dispersive, states~\cite{luttinger}, and a singular spectrum, with gaps of different sizes. 

{\bf Relaxation effects:}
For small twist angles $\theta$, the geometrical deformations of the moiré unit cell discussed here appear at very low strains of the order $\epsilon \sim \theta$. For the case of twisted bilayer graphene, the relaxation effects will shrink significantly the $AA$ regions, leading to the formation of one-dimensional channels, see for instance~\cite{Guinea2019,EM18,WG19,BDR20}, where the low energy electron states are defined. It can be expected that, at the critical strain in the minimal angle regime, the twisted bilayer graphene would be described by a network of parallel one-dimensional channels.

{\bf Conclusions:} We have presented a general geometry based approach to the strained bilayer graphene. It can be easily adopted to the larger class of strained and twisted bilayer systems. We have found simple expressions for the critical strain, at which the formation of one-dimensional strip-like moir\'e patterns occurs. We find that the formation of such patterns is a consequence of the interplay between twist and strain which gives rise to a collapsing of the reciprocal space unit cell.  The criterion for this transition appears to be a very simple relation between the applied uniaxial strain, the twist angle, and the material dependent Poisson ratio. Our results offer simple explanations for the complex patterns of one-dimensional channels observed in low angle twisted bilayer graphene systems and twisted bilayer dicalcogenides. 

The electronic bands in twisted bilayer graphene in the one dimensional regime are described by the interplay between two different, typically incommensurate, periodicities, suggesting similarities with the Harper equation and with one dimensional quasicrystals. 
%\textcolor{red}{we need to add something about the bands at criticality and Harper equations}

\vspace{5mm}
{\bf Acknowledgements:} 
A.S. was supported by the research grant PCI2021-122057-2B of the Agencia Estatal de Investigacion de Espa\~na. IMDEA Nanociencia acknowledges support from the \textquotedblleft Severo Ochoa\textquotedblright ~Programme for Centres of Excellence in R\&D (Grant No. SEV-2016-0686). P.A.P and F.G. acknowledge funding from the European Commission, within the Graphene Flagship, Core 3, grant number 881603 and from grants NMAT2D (Comunidad de Madrid, Spain), SprQuMat.

\bibliographystyle{apsrev4-1}

%merlin.mbs apsrev4-1.bst 2010-07-25 4.21a (PWD, AO, DPC) hacked
%Control: key (0)
%Control: author (72) initials jnrlst
%Control: editor formatted (1) identically to author
%Control: production of article title (-1) disabled
%Control: page (0) single
%Control: year (1) truncated
%Control: production of eprint (0) enabled
%

%\bibliography{references.bib}

\begin{thebibliography}{63}
\bibitem [{lut()}]{luttinger}%
  \BibitemOpen
  \href@noop {} {}\bibinfo {note} {It is worth noting that the velocity
  extracted from the bands in Fig.~\ref{fig:Figure4} is similar to the Fermi
  velocity in independent graphene layers. We will not consider here the role
  of interactions and the possibility that they may lead to Luttinger liquid
  behavior.}\BibitemShut {Stop}%
\end{thebibliography}

%\widetext
\clearpage
\onecolumngrid

\setcounter{equation}{0}
\setcounter{figure}{0}
\setcounter{table}{0}
\setcounter{page}{1}
\makeatletter
\renewcommand{\theequation}{S\arabic{equation}}
\renewcommand{\thefigure}{S\arabic{figure}}

\begin{center}
\Large Supplementary Materials for \\
Strain induced quasi-unidimensional channels in twisted moir\'e lattices \\
\vspace{5mm}
\normalsize Andreas Sinner, Pierre A. Pantale\'on and Francisco Guinea
\end{center}

\section{Strain induced quasi-unidimensional channels}
\begin{figure*}[h]
\includegraphics[scale=1.6]{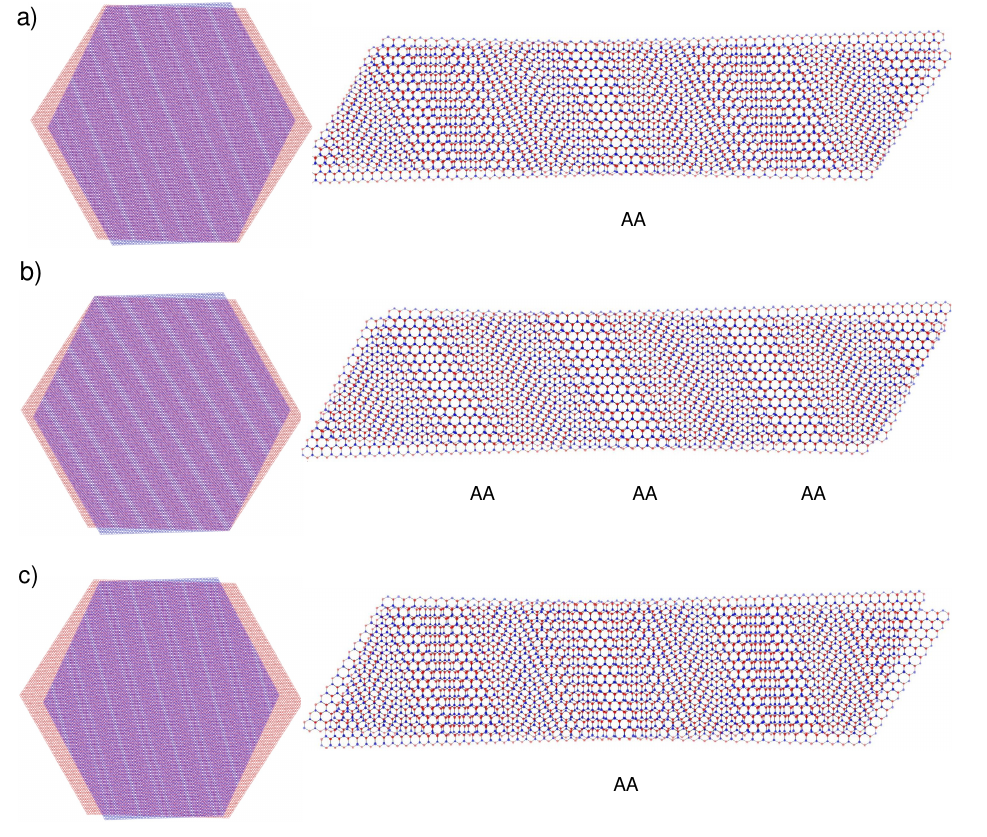}
\caption{
Strain induced quasi-unidimensional channels in the twisted bilayer moir\'e lattice. The twist angle is chosen at $3.0^\circ$ and the strain parameter is $x=1$ in all figures.  The ratio between the applied heterostrain is a) $- \epsilon_{yy} / \epsilon_{xx}$ = 0.165, b) $- \epsilon_{yy} / \epsilon_{xx}$ = 1/3 and c) $- \epsilon_{yy} / \epsilon_{xx}$ = 1/12. Figures on the right are enlarged cuts to visualize the stack configuration for each strain ratio. The AA stacking centers are indicated.  
}
\label{fig:RealCritical}
\end{figure*}

Figure~\ref{fig:RealCritical} shows a twisted bilayer moir\'e lattice under strain. The relative angle between honeycomb lattices is set to be $3^{\circ}$ for better visualization of the stacking and emergent patterns.  The stacking configuration is set to AA at the origin.  Figure~\ref{fig:RealCritical}a) shows the case of $- \epsilon_{yy} / \epsilon_{xx}$ = 0.165 (graphene Poisson ratio), b) of $- \epsilon_{yy} / \epsilon_{xx}$ = 1/3 and c) of $- \epsilon_{yy} / \epsilon_{xx}$ = 1/12. A commensurate periodicity reveals quasi-unidimensional channels with an $AA$ stack configuration. However, as explained in the main text, the relaxation of the atomic sites shrinks significantly these AA regions giving rise to unidimensional channels~\cite{Guinea2019}. 

\section{Critical strain and twist angle} 
\begin{figure*}[h]
\includegraphics[scale=1.0]{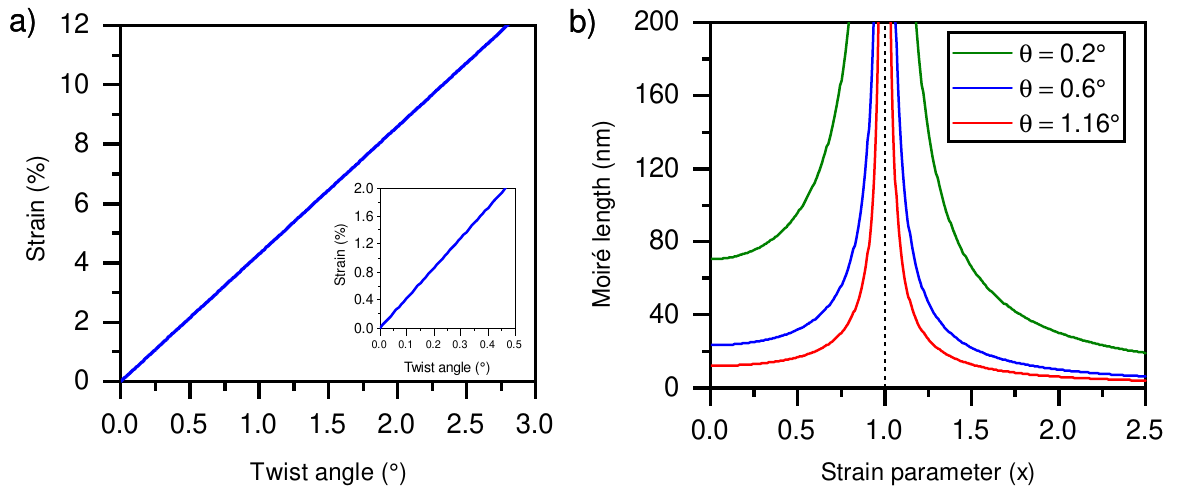}
\caption{
a) Critical strain as function of the twist angle, cf. Eq.~(\ref{eq:Collapse}) in the main text. For smaller twist angles (inset) the critical strain is within the experimental range. b) Moir\'e length versus strain parameter $x$ at different twist angles. At the critical point the moir\'e length diverges resulting in uni-dimensional channels in real space. Notice that after the critical point at $x=1$ is passed, the unit cell is again well defined. 
}
\label{fig:PlotsMoireAngle}
\end{figure*}

Figure~\ref{fig:PlotsMoireAngle}a) shows the critical strain percentage as a function of the twist angle, cf. Eq.~(\ref{eq:Collapse}) in the main text. As the twist angle is reduced, the strain required to form unidimensional channels becomes smaller. Figure~\ref{fig:PlotsMoireAngle}a) shows the magnitude of the moir\'e length as function of the strain parameter for different twist angles.

\section{Edge domain walls} 

\begin{figure*}[h]
\includegraphics[scale=1.5]{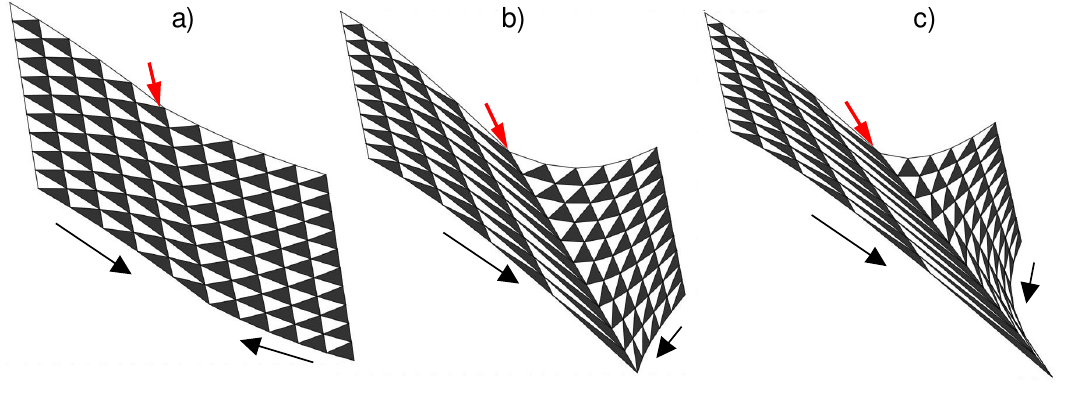}
\caption{
The visualization of the domain walls formation for different strain values. The non uniform strain increases from zero up to a finite value. At each maximum, there appears a domain wall indicated by  red arrows. For each panel this maximal strain is a) $0.3 \epsilon_{c}$, b)  $0.6 \epsilon_{c}$ and c)  $0.7 \epsilon_{c}$ 
}
\label{fig:DomainWalls}
\end{figure*}

Figure~\ref{fig:DomainWalls} shows different domain walls generated by the uniaxial heterostrain of different strength.  Black arrows indicate the direction of the strain increasing from zero to a finite value. Each triangle represents an AB/BA domain and the vertices in the triangles are at the position of the AA centers in each unit cell. These domains are constructed from the linear combinations of the lattice vectors defined in Eq.~(\ref{eq:RealSpaceV1}) and Eq.~(\ref{eq:RealSpaceV2}). 

\section{The continuum model of TBG} 
In the limit of sufficiently large periods of the moir\'{e} superlattices measured with respect to the atomic scale of the monolayer graphene, the physics of TBG is well described by the effective continuum model~ \cite{LopesdosSantos2007,Bistritzer2011,LopesdosSantos2012}. For strained twisted bilayer graphene the continuum model takes the form
\begin{equation}
\label{eq:HTBG}
H^{}_{\rm TBG} = \left(
\begin{array}{cc}
 H^{}({\bf q}^{\u}_\zeta)  & U^\dag(r) \\
U(r) & H^{}({\bf q}^{\d}_\zeta)
\end{array}
\right).
\end{equation}
The diagonal terms describe the Dirac particles in each layer 
\begin{equation}
H({\bf q}^{\ell}_\zeta) = -\hbar v^{}_F {\bf q}^{\ell}_\zeta \cdot (\zeta\sigma^{}_x,\sigma^{}_y),
\end{equation}
with $\zeta=\pm$ denoting the valley index, $\sigma^{}_{x,y}$ the Pauli matrices in the usual representation
and ${\bf q}^{\ell}_\zeta$ the momentum operator acting within the deformed moir\'e Brillouin zone 
\begin{equation}
{\bf q}^{\ell}_\zeta = R\left[\pm{\theta}/{2}\right]\left(\mathds 1 + {\cal E}^{\ell}\right)\left({\bf q} - {\bf D}^{\ell}_\zeta\right),
\end{equation}
with the rotation matrix $R[\pm\theta]$ 
\begin{equation}
\label{eq:RotMat}
R[\pm\theta] = 
\left(\begin{array}{cc} \cos\theta &  \mp\sin\theta \\ \pm\sin\theta & \cos\theta
\end{array}\right),
\end{equation}
and ${\bf q}=i{\bf\nabla}$.  The positions of the Dirac points in each layer are given by 
\begin{equation}
\label{eq:DiracPos}
{\bf D}^{\ell}_\zeta = \left(\mathds 1 - {\cal E}^{\ell}\right) R\left[\mp{\theta}/{2}\right]{\bf K}^{}_\zeta - \zeta {\bf A}^{\ell},
\end{equation}
where ${\bf K}^{}_\zeta = -\zeta/3\left( 2{\bf b}^{}_1 + {\bf b}^{}_2 \right) = {2\pi}\zeta({3a})^{-1} \left( -2 , 0\right)^{\rm T}$ with ${\bf b}^{}_{1,2}$ 
being the reciprocal lattice vectors of the monolayer graphene. The vector potential is defined as 
${\bf A}^{\ell} = {\sqrt{3}}\beta^{}_G({2a})^{-1} 
\left( \epsilon^{\ell}_{xx} - \epsilon^{\ell}_{yy} \;\; , -2\epsilon^{\ell}_{xy}\right)^{\rm T}, 
$
with the Gr\"uneisen dimensionless parameter $\beta^{}_G \approx 3.14$. For the experimentally relevant case of the uniaxial heterostrain along the $k^{}_x$-axis 
(i.e. for $\phi=0$ in Eq.(\ref{eq:uhs})), the expression for the vector potential simplifies to ${\bf A}^{\u,\d} = \mp{\bf A}$, where
\begin{equation}
\label{eq:VectPot}
{\bf A} = \frac{\sqrt{3}}{4a}\beta^{}_G\epsilon(1+\nu)
\left(
\begin{array}{cc}
1 \\ 0       
\end{array}
\right).
\end{equation}

The off-diagonal blocks in Eq.(\ref{eq:HTBG}) describe the interlayer coupling between twisted graphene layers in terms of the Fourier expansion,
\begin{equation}
\label{eq: Ucoupl}
U = 
U^{}_t + U^{}_l
e^{i\zeta\tilde{\bf g}^{}_1\cdot{\bf r}}
+
U^{}_r e^{i\zeta(\tilde{\bf g}^{}_1 + \tilde{\bf g}^{}_2)\cdot{\bf r}}, 
\end{equation}
with $\tilde{\bf g}^{}_i$ defined in Eqs.~(\ref{eq:StrBasis}), (\ref{eq:LatVec}), and  matrices 
$$
U^{}_t = \left(
\begin{array}{cc}
u & v \\ v & u       
\end{array}
\right),
\;\;
U^{}_l = \left(
\begin{array}{cc}
u & v\omega^{-\zeta} \\ v\omega^{\zeta} & u       
\end{array}
\right), 
\;\;
U^{}_r = \left(
\begin{array}{cc}
u & v\omega^{\zeta} \\ v\omega^{-\zeta} & u      
\end{array}
\right)
,
$$ 
with $\omega = \exp\left\{2\pi i/3\right\}$, $u = 0.0797$eV, and $ v = 0.0975$eV. To diagonalize the Hamiltonian in Eq.(\ref{eq:HTBG}) we restrict the number of wave vectors in Eq.~(\ref{eq:HTBG}). Without strain~\cite{Pantaleon2022}, the band structure can be accurately described with 91 wave vectors. However, as the strain increases, additional wave  vectors are needed. The wave vectors are simply given by the linear combination $n\tilde {\bf g}_1 + m\tilde {\bf g}_2$ with $n,m$ integers. The total number of vectors is set to be large enough to achieve convergence.

\section{The reciprocal lattice vectors at the critical strain} 

From Eqs.~(\ref{eq:LatVec}) and Eq.~(\ref{eq:CollapseEq}),  we get the reciprocal lattice vectors at the critical strain,
\begin{eqnarray}
\tilde {\bf g}^{}_{1,2} &=& \left(
\mp\sqrt{3} - \frac{3}{\sqrt{\nu}} 
\right) \tilde k^{}_0
\left(
\begin{array}{c}
 1
\\
\sqrt{\nu} 
\end{array}
\right).
\end{eqnarray}
Then, the linear combination of both vectors yields
\begin{equation}
n\tilde {\bf g}_1 + m\tilde {\bf g}_2 = \sqrt{3}\tilde k^{}_0
\left[m - n - \sqrt{\frac{3}{\nu}} (n+m)\right]
\left(
\begin{array}{c}
 1
\\
\sqrt{\nu} 
\end{array}
\right).
\label{eq: lincomb}
\end{equation}
Obviously, the single periodicity condition requires 
\begin{equation}
n\tilde {\bf g}_1 + m\tilde {\bf g}_2 = 0,
\end{equation}
or correspondingly
\begin{equation}
\frac{1}{\sqrt{\nu}} = \frac{1}{\sqrt{3}} \frac{m-n}{m+n},
\label{eq:periodcond}
\end{equation}
from which Eq.~(\ref{eq:PoisonInte}) follows.

\section{Evolution of the TBG band structure with strain}

\begin{figure}[h]
\includegraphics[scale=0.7]{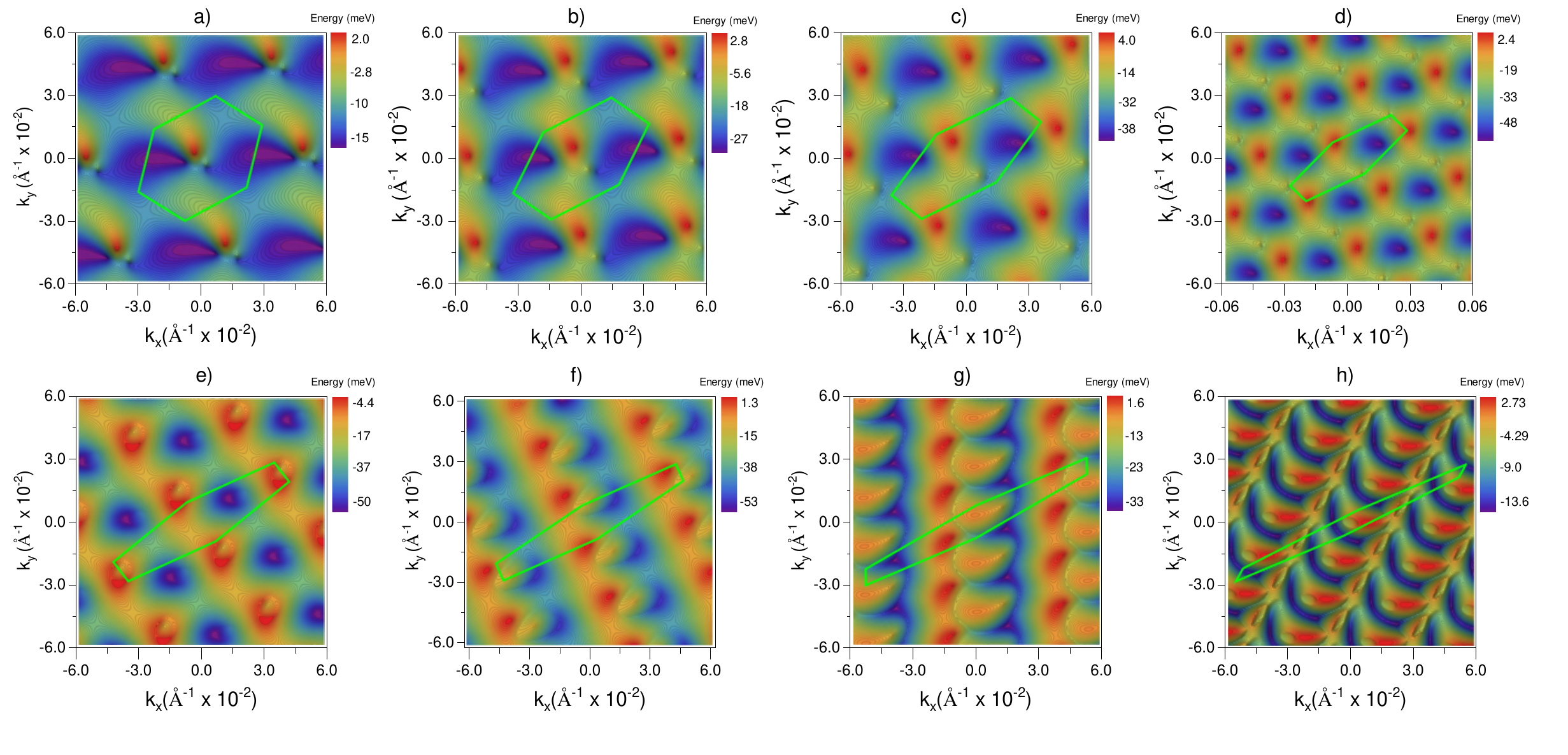}
\caption{Evolution of the bottom middle band of TBG with the twist angle $\theta=1.0^{\circ}$ as  function of the strain for: a) $0.1 \epsilon_{c}$, b) $0.2 \epsilon_{c}$, c) $0.3 \epsilon_{c}$, d) $0.4 \epsilon_{c}$, e) $0.5 \epsilon_{c}$ f) $0.6 \epsilon_{c}$, g) $0.7 \epsilon_{c}$ and h) $0.8 \epsilon_{c}$;  $\epsilon_{c}$ denotes the critical strain value defined in Eq.~(\ref{eq:Collapse}). Green hexagon emphasizes the moir{\'e} Brillouin zone.
}
\label{fig:Bandsevol}
\end{figure}

\begin{figure*}[h]
\includegraphics[scale=0.4]{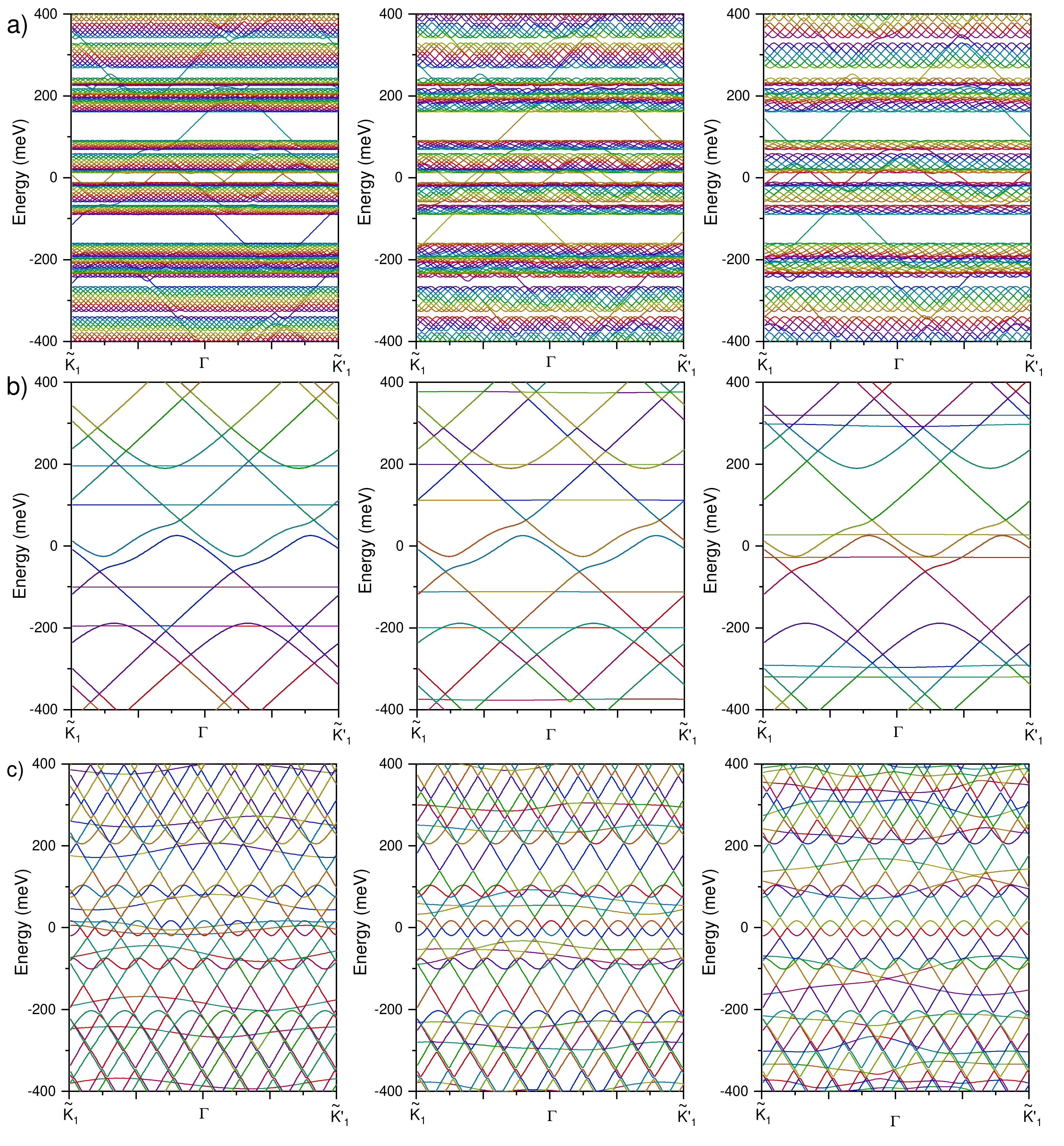}
\caption{Band structures of the critically strained TBG for $\theta = 1^{\circ}$ and different values of the ratio between applied strains, from top to bottom: a) $- \epsilon_{yy} / \epsilon_{xx}$ = 0.165, b) $- \epsilon_{yy} / \epsilon_{xx}$ = 1/3 and c) $- \epsilon_{yy} / \epsilon_{xx}$ = 1/12. The spectra are evaluated along the collapsed Brillouin zone depicted in Fig.~\ref{fig:Figure1}. The columns are calculated with decreasing precision  (from left to right) for the Hamiltonian Eqs.~(\ref{eq:HTBG})-(\ref{eq: Ucoupl}) truncated at $1951$, $1261$, and $721$ $\bf\tilde g$-vectors in the reciprocal space. The convergence is almost reached in both cases with a single periodicity (middle and bottom rows). The spectrum of the $- \epsilon_{yy} / \epsilon_{xx}$ = 0.165 realization (top row) reveals new structures at every higher precision, typical for aperiodical systems. 
}
\label{fig:ComparisonCritical}
\end{figure*}

Figure~\ref{fig:Bandsevol} shows the evolution of the lower middle band of the twisted bilayer graphene as  function of the strain. With  increasing strain values, the cones move on complex trajectories thorough the moir{\'e} Brillouin zone~\cite{Fu2019,Pantaleon2022,Pantaleon2021,Zhang2022Strain}. The complexity of this trajectory is due to the interplay between energetic and geometric contributions, cf. supplementary materials in Ref.~\cite{Pantaleon2022}. 
At the critical strain, the moir{\'e} Brillouin zone collapses to a line and the wave vectors become parallel to each other following Eq.~(\ref{eq: lincomb}). For numerical diagonalization one needs to truncate the Hamiltonian at some finite size. The spectra obtained with different precision are shown in Fig.~\ref{fig:ComparisonCritical}. 

\end{document}